\documentclass[twocolumn,aps]{revtex4}
\usepackage{mathrsfs}
\usepackage{}

\pdfoutput=1

\usepackage[english]{babel}

\usepackage{dcolumn}

\usepackage{amsmath}
\usepackage{graphicx}
\usepackage{subfigure}
\usepackage{multirow}
\usepackage{txfonts}
\def\bea{\begin{eqnarray}}
\def\eea{\end{eqnarray}}
\def\be{\begin{equation}}
\def\ee{\end{equation}}
\def\qu{quintessence }

\begin{document}

\title{Observational constraint on the dark energy scalar field }

\author{Ming-Jian Zhang$^{a}$}
\author{Hong Li$^{a}$}\email[Corresponding author: ]{hongli@ihep.ac.cn}
\affiliation{$^a$Key Laboratory of Particle Astrophysics, Institute of High Energy Physics,
Chinese Academy of Science, P. O. Box 918-3, Beijing 100049, China}

\begin{abstract}
In the present paper, we investigate three scalar fields, \qu field, phantom field and tachyon field, to explore the source of dark energy, using the Gaussian processes method from the background data and perturbation growth rate data. The corresponding reconstructions all suggest that the dark energy should be dynamical. Moreover, the quintom field, a combination between \qu field and phantom field, is powerfully favored by the data within 68\% confidence level. Using the mean values of scalar field $\phi$ and potential $V$, we fit the function $V(\phi)$ in different fields. The fitted results imply that potential $V(\phi)$ in each scalar field may be a double exponential function or Gaussian function. The Gaussian processes reconstructions also indicate that the tachyon scalar field cannot be convincingly favored by the data and is at a disadvantage to describe the dark energy.

\end{abstract}

%\pacs{95.36.+x, 98.80.-k, 98.80.Es, 98.80.Jk}
\maketitle

\section{Introduction}
\label{introduction}

Multiple experiments, including the type Ia supernova (SNIa) \citep{riess1998supernova,perlmutter1999measurements}, large scale structure \citep{tegmark2004cosmological}, cosmic microwave background (CMB)
anisotropies  \citep{spergel2003wmap}, and baryon
acoustic oscillation (BAO) peaks \citep{eisenstein2005detection} have consistently approved that our universe was accelerating expansion. Theoretically, this acceleration needs a new component with repulsive gravity to drive. In numerous theoretical paradigms, the exotic dark energy theory focused on the most attention. One essential parameter to understand the nature of dark energy, is the equation of state (EoS) $w$, the ratio of pressure to energy density. In recent analysis \cite{ade2016planck}, the cosmological constant model with  $w=-1$ fits well with the Planck data and other astrophysical data. However, many other data also mildly favour an evolving dark energy, especially in the very recent extended BAO survey \cite{zhao2017dynamical}. Scalar field theory, such as the \qu field \cite{wetterich1988cosmology,wetterich1995asymptotically,ratra1988cosmological,caldwell1998cosmological}, phantom field \cite{caldwell2002phantom}, tachyon field \cite{sen2002rolling,sen2002tachyon,sen2002field}, is one program to achieve the evolving dark energy. For the \qu field, it has a positive kinetic energy density with $-1 \leq w \leq 1$. While for the phantom field, it has a negative kinetic energy density with $ w \leq -1$.

However, Refs. \cite{feng2005dark,zhao2005perturbations,caldwell2005dark}  proved that the dark energy perturbation would be divergent when  $w$ approaches to $-1$, in the \qu and phantom models.
Moreover, numerous observations favor a $w$ crossing $-1$ in the near past. Regretfully, neither the
\qu nor phantom scalar field can fulfill this transition. To solve these problems, \citet{feng2005dark} proposed the quintom model, a combination of \qu field $\phi_1$ and phantom field $\phi_2$ in the Lagrangian. When the time derivative of scalar field $\dot{\phi}_1 > \dot{\phi}_2$, it leads to $ w \geq -1$; while for $\dot{\phi}_1 < \dot{\phi}_2$, we have $ w \leq -1$. To promote the quintom model being a single scalar field, Refs. \cite{li2005single,guo2005cosmological,zhang2006avoiding,apostolopoulos2006late,li2012dark} introduce higher derivative operators in the Lagrangian. They found that the models are consistent with the observations. The $w$ also can cross $-1$ without any instability.

However, we notice that the potentials $V(\phi)$ in the literatures were mostly built by a parametrization of scalar field, either the \qu, phantom or the quintom models. The common popular templates were the power-law potential, exponential potential, or trigonometric function potential, and so on. However, a $V(\phi)$ template inevitably imposes a strong prior on the underlying property of cosmic dynamics. In our view, a straightforward manner and template-free study has an advantage to understand the cosmic dynamics.

In this paper, we focus on a prominent technique, the Gaussian processes (GP) analysis. Unlike the parametrization constraint, this approach does not rely on any artificial potential template. It can be actualized via a purely statistical manner. In this process, it firstly assumes that each observational data satisfies a Gaussian distribution. Thus, the data sets should satisfy a multivariate normal distribution. Any two different data points are connected by a covariance function $k(z, \tilde{z})$. Using the function $k(z, \tilde{z})$, information of the observational variable can be extrapolated to other redshift which have not be observed. Finally, with this information, an involved goal function, potential $V$ can be reconstructed. We note that the primary task in this Gaussian processes is to determine the function $k(z, \tilde{z})$ using the observational data. Because this process is independent of any template of the goal function, this approach has incurred a wide application in cosmology  \cite{seikel2012reconstruction,seikel2012using,yahya2014null,zhang2016test,wang2017improved,wei2017improved,zhang2018physical,haridasu2018improved,pinho2018model,sangwan2018reconstructing}. In our recent work \cite{zhang2018gaussian}, we investigated the dark energy using this method. We found that the background and perturbation data both present a hint of dynamical dark energy. However, a further understanding on the dynamics of these reconstructed $w$ is absent. In this paper, we would like to provide a further analysis from the scalar field. Namely, the goal of this work is to explore which scalar field may be the dynamical source of dark energy. This test can update our understanding on the cosmic acceleration by presenting a model-independent result. Following our recent work, we still focus on the background data from supernova and Hubble parameter, and perturbation data from the growth rate of structure. The scalar fields we consider are \qu, phantom, and tachyon.

This paper is organized as follows: In Section~\ref{theory}, we introduce the scalar field and GP approach. And in Section~\ref{data} we introduce the relevant data we use. We present the reconstruction result in Section \ref{result}. Finally, in Section \ref{conclusion} conclusion and discussion are drawn.

\section{Theory and method}
\label{theory}

In this section, we introduce some theoretical basis about the scalar field and GP approach.

\subsection{Scalar field}  \label{field}

For a spatial flat Friedmann-Robertson-Walker universe with dark matter and scalar field, the Friedman equations are
\bea  \label{Friedman_equation}
H^2 &=& \frac{8\pi G}{3} (\rho_m + \rho_{\phi} ),  \nonumber \\
\frac{\ddot{a}}{a} &=& -\frac{4\pi G}{3} (\rho_m + \rho_{\phi} + 3 p_{\phi}) ,
\eea
where the Hubble expansion rate $H=\dot{a}/a$ is a function of scalar factor $a(t)$, and the dot denotes derivative with respect to cosmic time $t$. The parameter $\rho_{\phi}$ and $p_{\phi}$ are energy density and pressure of scalar field, respectively. For the dark matter, its energy density yields $\rho_m = \rho_{m0} (1+z)^3$, where $\rho_{m0}$ is the current energy density. Generally, we introduce the energy density parameter $\Omega_{m0} = \rho_{m0}/\rho_{c0}$, with critical density $\rho_{c0}=3H_0^2/(8 \pi G)$, where $H_0$ is the Hubble constant. For the scalar filed, we consider three scenarios in this present paper, namely, the \qu, phantom and tachyon scalar field.

For the \qu scalar field, its energy density and pressure are defined as
\bea \label{def:\qu}
\rho_{\phi} &=& \frac{1}{2}\dot{\phi}^2 + V(\phi) ,  \nonumber \\
p_{\phi} &=& \frac{1}{2}\dot{\phi}^2 - V(\phi) .
\eea
For the potential $V(\phi)$, many models were proposed (see Ref. \cite{tsujikawa2013quintessence} for a short review ), such as the  power-law potential $V(\phi) \propto \phi^{p}$, exponential potential $V(\phi) \propto e^{-\lambdaup \phi}$, inverse power-law potential $V(\phi) \propto \phi^{-p}$, inverse exponential potential $V(\phi) \propto e^{\lambdaup/ \phi}$, double exponential potential $V(\phi) \propto V_1 e^{-\lambdaup_1 \phi} + V_2 e^{-\lambdaup_2 \phi}$, Hilltop potential $V(\phi) \propto \cos(\phi)$. Meanwhile, some other complex models also can be found in Ref. \cite{sahni20045}, such as $e^{\lambdaup \phi^2}/\phi^{\alpha}$, $(\cosh \lambdaup \phi -1)^p$, $\sinh ^{-\alpha} (\lambdaup \phi)$, $[(\phi - B)^{\alpha} + A] e^{-\lambdaup \phi}$. Recently, some of these models were constrained using the observational data \cite{wang2012confronting,takeuchi2014probing,mukherjee2015reconstruction,linder2015quintessence,sangwan2018observational}. They found that some models cannot be discriminated, or some ones are not disfavored by the observational data, such as the inverse power-law potential, inverse exponential potential, even some ones have intrinsic limitation. Now, we turn to the solution of the \qu scalar field. Putting the definition of Eq. \eqref{def:\qu} to Friedman equations \eqref{Friedman_equation}, we can solve the \qu field and potential as
\bea \label{\qu:phi2t}
\frac{8\pi G}{3H_{0}^2} \dot{\phi}^{2} &=& \frac{1}{3}(1+z) E^{2 \prime}  - \Omega_{m0} (1+z)^3 , \nonumber\\
\frac{8\pi G}{3H_{0}^2} V &=& E^2 - \frac{1}{6}(1+z) E^{2 \prime} - \frac{1}{2}\Omega_{m0} (1+z)^3 ,
\eea
where the prime denotes derivative with respect to redshift $z$; $E(z) = H(z)/H_0$ is the dimensionless Hubble parameter. We find that, on the one hand, both the derivative of scalar field $\dot{\phi}^{2}$ and potential $V$ are in units of $\frac{8\pi G}{3H_{0}^2}$. On the other hand, the function $\dot{\phi}^{2}$ may be negative when the former term is less than the latter term. If this case happens, it would change into the other model, the phantom scalar field.

For the phantom scalar field, its energy density and pressure are
\bea  \label{def:phantom}
\rho_{\phi} &=& -\frac{1}{2}\dot{\phi}^2 + V(\phi) , \nonumber \\
p_{\phi} &=& -\frac{1}{2}\dot{\phi}^2 - V(\phi) .
\eea
Performing a similar calculation, we can obtain the phantom field and potential
\bea  \label{phantom:phi2t}
\frac{8\pi G}{3H_{0}^2} \dot{\phi}^{2} &=& \Omega_{m0} (1+z)^3 - \frac{1}{3}(1+z) E^{2 \prime} ,   \nonumber\\
\frac{8\pi G}{3H_{0}^2} V &=& E^2 - \frac{1}{6}(1+z) E^{2 \prime} - \frac{1}{2}\Omega_{m0} (1+z)^3 .
\eea
Obviously, function $\dot{\phi}^{2}$ in the Eq. \eqref{phantom:phi2t} is opposite to the $\dot{\phi}^{2}$ in Eq. \eqref{\qu:phi2t}. Therefore, the \qu and phantom field, no more than one model can survive. Similar to above \qu scalar field, the cosmologist also modelled a lot of phantom potentials. \citet{PhysRevLett.91.071301} studied this scalar field, and found that $w<-1$ would cause a big rip of the universe. Investigation in Ref. \cite{roy2018dynamical} considered five models, and showed that they fit well with the observational data, but no one occupies a special position. In order to solve the problem of  $w$ crossing $-1$ in the near past from $w > -1$ to $w < -1$, \citet{feng2005dark} proposed the quintom model, a combination of \qu field and phantom field in the Lagrangian with a double exponential potential
\bea  \label{L:quintom}
\mathcal{L} &=& \frac{1}{2} \partial_{\mu} \phi_1 \partial^{\mu} \phi_1 - \frac{1}{2} \partial_{\mu} \phi_2 \partial^{\mu} \phi_2 \nonumber\\
              &&- V_0 \big[\exp(-\frac{\lambdaup}{m_p} \phi_1) + \exp(-\frac{\lambdaup}{m_p} \phi_2) \big] .
\eea
They found that this model also satisfies the observations.

For the tachyon scalar field, it is a different scalar field from above two scenarios. Its energy density and pressure are
\bea  \label{def:tachyon}
\rho_{\phi} &=& \frac{V(\phi)}{\sqrt{1-\dot{\phi}^2}} ,   \nonumber \\
p_{\phi} &=& -V(\phi) \sqrt{1-\dot{\phi}^2} .
\eea
Combining with the Friedman equations \eqref{Friedman_equation}, the tachyon field and potential can be solved as
\bea  \label{tachyon:phi2t}
\dot{\phi}^{2} &=& \frac{(1+z) E^{2 \prime} - 3\Omega_{m0} (1+z)^3}{ 3E^2 - 3\Omega_{m0} (1+z)^3}     ,  \nonumber\\
\frac{8\pi G}{3H_{0}^2} V &=&  \sqrt{E^2-\frac{1+z}{3}  E^{2 \prime}} \sqrt{E^2 - \Omega_{m0} (1+z)^3} .
\eea
For this solution, we have several points to note. Firstly, the term $1-\dot{\phi}^2$ in Eq. \eqref{def:tachyon} must be positive. Secondly, we find that the solutions  $\dot{\phi}^{2}$ and $V$ in Eq. \eqref{tachyon:phi2t} are much different from the ones in above two scenarios. For the function $\dot{\phi}^{2}$, it is immune from the nuisance parameter $\frac{8\pi G}{3H_{0}^2}$. For the potential $V$, it should be non-negative in the square root of Eq. \eqref{tachyon:phi2t}. In cosmology, several tachyon models were studied. In Ref. \cite{calcagni2006tachyon}, the authors numerically investigated a range of potentials, and found that tachyon models have quite similar phenomenology to canonical \qu models. And also, some models are not strongly disfavoured by observations \cite{barbosa2018theoretical}. However, Ref. \cite{guo2004cosmological} found that the universe could accelerate only at nearly Planck energy densities, for a single tachyon field with an inverse square potential. The acceleration should be driven by multiple tachyon fields at
lower-Planck energy densities.

To obtain the potential $V(\phi)$, we should solve the field $\phi$ from the function $\dot{\phi}^{2}$. Using the relation $dt = -\frac{1}{(1+z)H} dz$, we can transfer the derivative of scalar field $\dot{\phi}^{2}$ over time $t$ to redshift $z$, namely,
\be
\left( \frac{d\phi}{dz} \right)^2 = \frac{\dot{\phi}^{2}}{(1+z)^2 H^2} .
\ee
Here we should be careful for the units of function $\left( \frac{d\phi}{dz} \right)^2$ in different scenarios. In our calculation, we reduce it to a dimensionless quantity. To obtain a dimensionless one, the function $\dot{\phi}^2$ in \qu and phantom fields, and potential $V(z)$ in these three fields are in units of $\frac{8\pi G}{3H_{0}^2}$. Function $\frac{d\phi}{dz} $ is in units of $H_0$. Theoretically, the function $\frac{d\phi}{dz}$ can take two symbols. Here we consider the positive values. Finally, the scalar filed can be obtained by
\be
\phi =   \int \frac{d\phi}{dz} dz .
\ee
In our calculation, we take the initial value $\phi_0 = 0$. After above preparations, the dimensionless potential $V$ and scalar field $\phi$ can be reconstructed. Thus, the function $V(\phi)$ can be modelled via a model-independent way in the following context.

\subsection{Methodology}
\label{methodology}

In the present paper, the data we use are background data from supernova and Hubble parameter; and perturbation data from redshift-space distortions (RSD).

For the background data, the distance modulus in the Friedmann-Robertson-Walker universe is
\begin{equation}  \label{mu:define}
    \mu (z) =  5 \textrm{log}_{10}d_L(z)+25,
\end{equation}
with the luminosity distance function
\begin{equation}
    \label{dL:define}
    d_L(z) =  \frac{c}{H_0} (1+z)  \int^z_0 \frac{  \mathrm{d}
    \tilde{z}}{E(\tilde{z})} .
\end{equation}
By introducing a dimensionless comoving luminosity distance
\be   \label{D:define}
D(z) \equiv \frac{H_0}{c} \frac{d_L (z)}{1+z} ,
\ee
we can obtain the relation between Hubble parameter and distance $D(z)$ via the Eqs. \eqref{D:define} and \eqref{dL:define}
\be  \label{Hubble_SN}
E (z)  = \frac{ 1}{D'} .
\ee

For the perturbation data, we consider a background universe filled with dark matter and unclustered dark energy scalar field. The evolution of matter density contrast, $\delta(z) \equiv \frac{\delta \rho_m}{\rho_m} (z)$, at scales much smaller than the Hubble radius should obey the following second order differential equation
\begin{equation}  \label{eq:delta}
   \ddot{\delta} + 2H \dot{\delta} -4\pi G \rho_m \delta =0 ,
\end{equation}
where $\rho_m$ is the background matter density, $\delta \rho_m$ represents its first-order
perturbation. According to the relation between scale factor and redshift, Hubble parameter in Eq. \eqref{eq:delta} can be expressed as an integral over the perturbation and its derivative  \cite{starobinsky1998determine,chiba2007consistency}
\be  \label{solution:E_delta}
E^2(z) = 3 \Omega_{m0} \frac{(1+z)^2}{\delta '(z) ^2 } \int_z^{\infty} \frac{\delta}{1+z} (-\delta ') d z .
\ee
We find that the Hubble parameter $E^2(z)$ tends to zero when the redshift in integral $z \rightarrow \infty$. When the redshift $z=0$, we have the initial condition
\be  \label{initial_condition}
1 = \frac{3 \Omega_{m0}}{\delta '(z=0) ^2} \int_0^{\infty} \frac{\delta}{1+z} (-\delta ') d z .
\ee
Using this initial condition, we consequently rewrite the Hubble parameter in Eq. \eqref{solution:E_delta}  as
\be  \label{Hubble_delta}
E^2(z) = (1+z)^2  \frac{\delta '(z=0) ^2}{\delta '(z) ^2 } \left[ 1 -  \frac{\int_0^{z} \frac{\delta}{1+z} (-\delta ') d z}{\int_0^{\infty} \frac{\delta}{1+z} (-\delta ') d z}  \right]  .
\ee

Observationally, current cosmological surveys cannot provide direct measurement of perturbation $\delta(z)$, but can provide a related observation, the growth rate measurement $f\sigma_8$ from  RSD. Here, the growth rate $f$ is defined by the derivative of the logarithm of perturbation $\delta$ with respect to logarithm of the cosmic scale
\be  \label{f:define}
f \equiv \frac{d \, \textmd{ln} \delta}{d \, \textmd{ln} a}  = -(1+z) \frac{d \, \textmd{ln} \delta}{d \, z} = -(1+z) \frac{\delta '}{\delta} .
\ee
While the function
\be
\sigma_8 (z) = \sigma_8 (z=0) \frac{\delta (z)}{\delta (z=0)}
\ee
is the linear theory root-mean-square mass fluctuation within a sphere of radius $8h^{-1}$Mpc. In the light of above two definitions, the growth rate of structure is written as
\be
f \sigma_8 = -\frac{\sigma_8 (z=0)}{\delta (z=0)} (1+z) \delta ' .
\ee
It is easy for us to have
\be  \label{Eq:dp}
\delta ' = - \frac{\delta (z=0)}{\sigma_8 (z=0)} \frac{f \sigma_8}{1+z}  .
\ee
Obviously, derivative of the perturbation $\delta$ can be easily transferred or reconstructed from the observational RSD data. Taking an integral to the two sides of Eq. \eqref{Eq:dp} over redshift, we have
\be  \label{delta_end}
\delta  = \delta (z=0) - \frac{\delta (z=0)}{\sigma_8 (z=0)} \int_0^{z} \frac{f \sigma_8}{1+z} dz .
\ee
For the constant $\delta (z=0)$, it was commonly considered as the normalization value $\delta (z=0)=1$ \cite{zhang2018gaussian}. For the other constant, we consider it as   $\sigma_8 (z=0)=0.8159$ \cite{ade2016planck}.

With above preparation of the theory, we can reconstruct the goal function $f(z)$ using the GP method. For the parametrization constraint, a prior template on the constrained function $f(z)$ is usually restricted. Different from it, the model-independent GP method, is not enslaved to any particular parametrization form. It only needs a probability on the goal function $f(z)$.  Assuming each observational data, such as the distance $D$, obeys a Gaussian distribution with mean and variance, the posterior distribution of all observed distance $D$ would obey the joint Gaussian distribution. In this process, the key ingredient is the covariance function $k(z, \tilde{z})$ which correlates the values of different distance $D(z)$ at points $z$ and $\tilde{z}$. Commonly, the covariance function $k(z, \tilde{z})$ has several types, and most is associated with two hyperparameters $\sigma_f$ and $\ell$ which can be determined by the observational data via a marginal likelihood. With the trained covariance function, the data can be extended to more redshift points. Using the relation between the goal function $f(x)$ and distance $D$, the former can be reconstructed. Due to its model-independence, this method has been widely applied in the reconstruction of dark energy EoS \cite{seikel2012reconstruction}, or in the test of the concordance model \cite{seikel2012using,yahya2014null}.

For the covariance function $k(z, \tilde{z})$, many forms are available. In the present paper, we adopt the most commonly used squared exponential
\begin{eqnarray}
k(z, \tilde{z}) = \sigma_f^2 \exp \left[\frac{-|z-\tilde{z}|^2}{2 \ell^2}  \right] .
\end{eqnarray}
With the chosen covariance function, we can reconstruct the scalar field by modifying the publicly available package GaPP  \cite{seikel2012reconstruction}. We also refer the reader to Ref. \cite{seikel2012reconstruction} for more details on the GP method.

\section{Observational data}
\label{data}

In this section, we report the related observational data.

For the supernova data, we use the joint light-curve analysis (JLA) datasets from the SDSS-II and SNLS surveys \cite{2014A&A...568A..22B}. Usually, they are presented as tabulated distance modulus with errors. For these JLA samples, they span a wide range at redshift $0.01 <z <1.3$. It consists of 740 SNIa datasets, including three-season data from SDSS-II ($0.05 < z <0.4$), three-year data from SNLS ($0.2 < z <1$), HST data ($0.8 < z <1.4$), and several low-redshift samples ($z <0.1$). According to their test, the binned JLA data have a same constraint power as the full version of the JLA likelihood on the cosmological model. In our calculation, we use the 31 binned distance modulus with covariance matrix, which is issued in their Ref. \cite{2014A&A...568A..22B}. In our calculation, we set the same prior of $H_0$ as the following $H(z)$ data. Moreover, the theoretical initial conditions $D(z=0)=0$ and $D'(z=0)=1$ are also taken into account in the calculation.

For the $H(z)$ data, they were not direct products from a tailored telescope, but can be acquired via two ways. One is to calculate the differential ages of galaxies
\cite{jimenez2008constraining,simon2005constraints,stern2010cosmic}, usually called cosmic chronometer. The other is the deduction from the BAO peaks in the galaxy power spectrum
\cite{gaztanaga2009clustering,moresco2012improved} or from the BAO
peak using the Ly$\alpha$ forest of QSOs \cite{delubac2013baryon}. In the present paper, we use the 30 cosmic chronometer data points which were compiled in our recent work \cite{zhang2016test}. For the latter method, it is model-dependent because an underlying cosmology is needed to calculate the sound horizon. Considering the error of Hubble constant, we can calculate the uncertainty of $E(z)$
\be
\sigma_E^2 = \frac{\sigma_H^2}{H_0^2} + \frac{H^2}{H_0^4} \sigma_{H_0}^2 .
\ee
We utilize the same prior of $H_0$ as the supernova data. Different from previous most work, we do not use the $H(z)$ data alone. We combine them with the supernova data as a derivative of distance $D$, using the relation
$
D' = \frac{1}{E(z)}.
$
Moreover, the initial condition $E(z=0)=1$ should be taken into account in our calculation. To test the influence of Hubble constant on the reconstruction,  we respectively consider two priors, namely, $H_0=73.24 \pm 1.74$ km s$^{-1}$Mpc$^{-1}$ with 2.4\% uncertainty \cite{riess20162} and  $H_0=67.74 \pm 0.46$ km s$^{-1}$Mpc$^{-1}$ from the latest determination \cite{ade2016planck}.

For the RSD data, they are in fact effects due to the differences between the observed distance and true distance on the galaxy distribution in redshift space. These differences are caused by the velocities in the overdensities deviation from the cosmic smooth Hubble flow expansion. Anisotropy of the radial direction  relative to transverse direction in the clustering of galaxies is correlated with the cosmic structure growth. Smaller deviation from the General Relativity implies a smaller anisotropic distortion in the redshift space. Based on above advantages, the RSD data is a very promising probe to distinguish the cosmological models, because different cosmological models may have similar background evolution, but the growth of structure may be very distinct. Till now, the RSD data have been used extensively in previous literatures.  In this paper, we utilize the most recent RSD data from 2dF, 6dF, BOSS, GAMA, WiggleZ, eBOSS DR14, galaxy surveys. We collected the compilation in our recent work \cite{zhang2018gaussian}, which includes the survey, RSD data with errors, the corresponding references and year.

\section{Result}
\label{result}

From above scalar fields and potentials, we find that their determinations are dependent of the Hubble parameter and matter density parameter. For the matter density parameter, we consider a moderate estimation $\Omega_{m0}=  0.279 \pm 0.025$ \cite{hinshaw2012nine}.  We know very well that the function $V(\phi)$ has always been the goal we've been pursuing. In the past few decades, a lot of models have been proposed, as introduced in above section. In this paper, we first reconstruct the scalar field $\phi (z)$ and potential $V(z)$; then we try to fit the function $V(\phi)$ using their mean values. Due to the model-independence of GP method, we think that it can give a more scientific test.

\subsection{Reconstruction from the JLA and $H(z)$ data}
\label{result_JLAOHD}

\begin{figure}
\includegraphics[width=9.2cm,height=5cm]{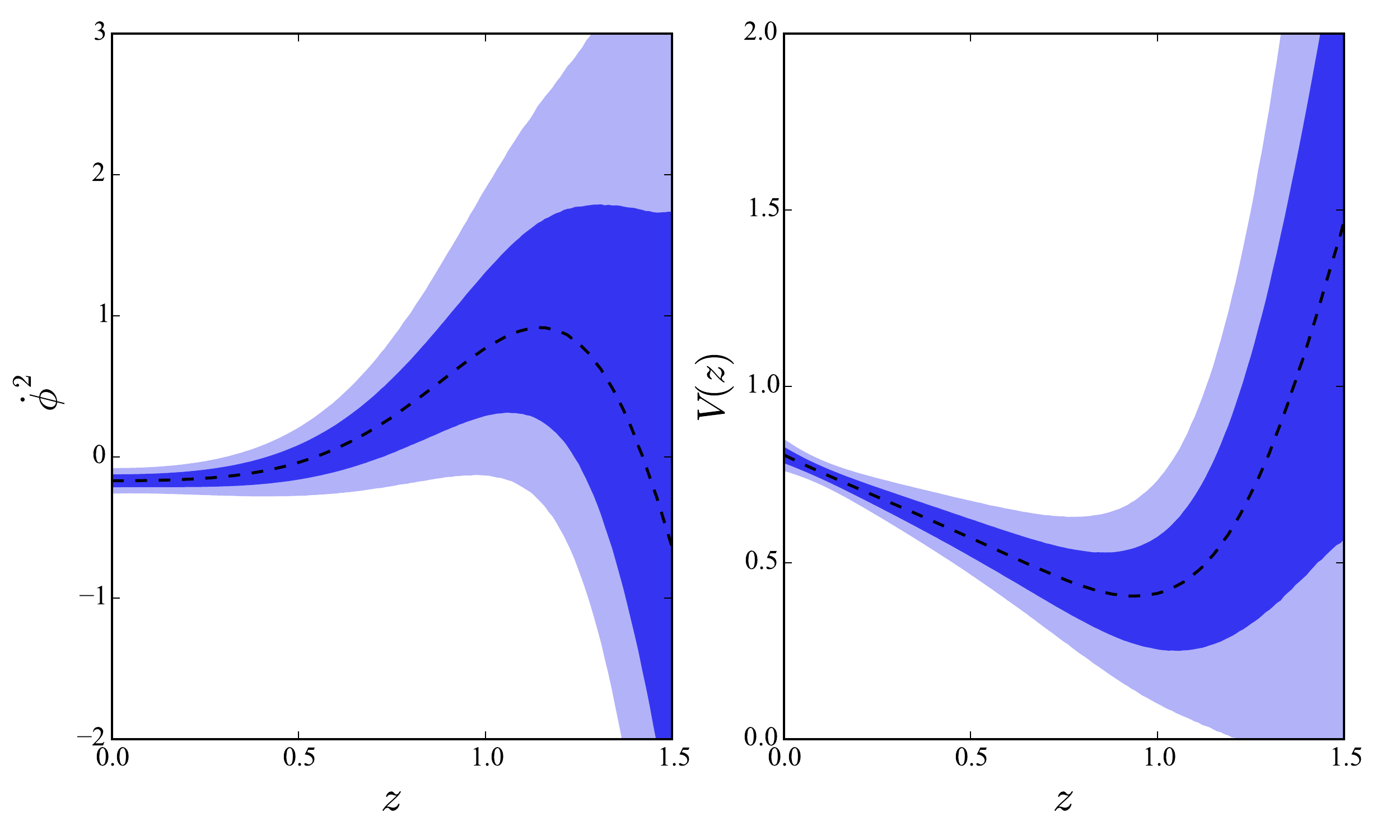}
    \caption{\label{JLAOHD_H07324_quintess} GP reconstruction in the \qu field for JLA and $H(z)$ data with Hubble constant $H_0=73.24 \pm 1.74$ km s$^{-1}$Mpc$^{-1}$. }
\end{figure}

\begin{figure}
\includegraphics[width=9.2cm,height=5cm]{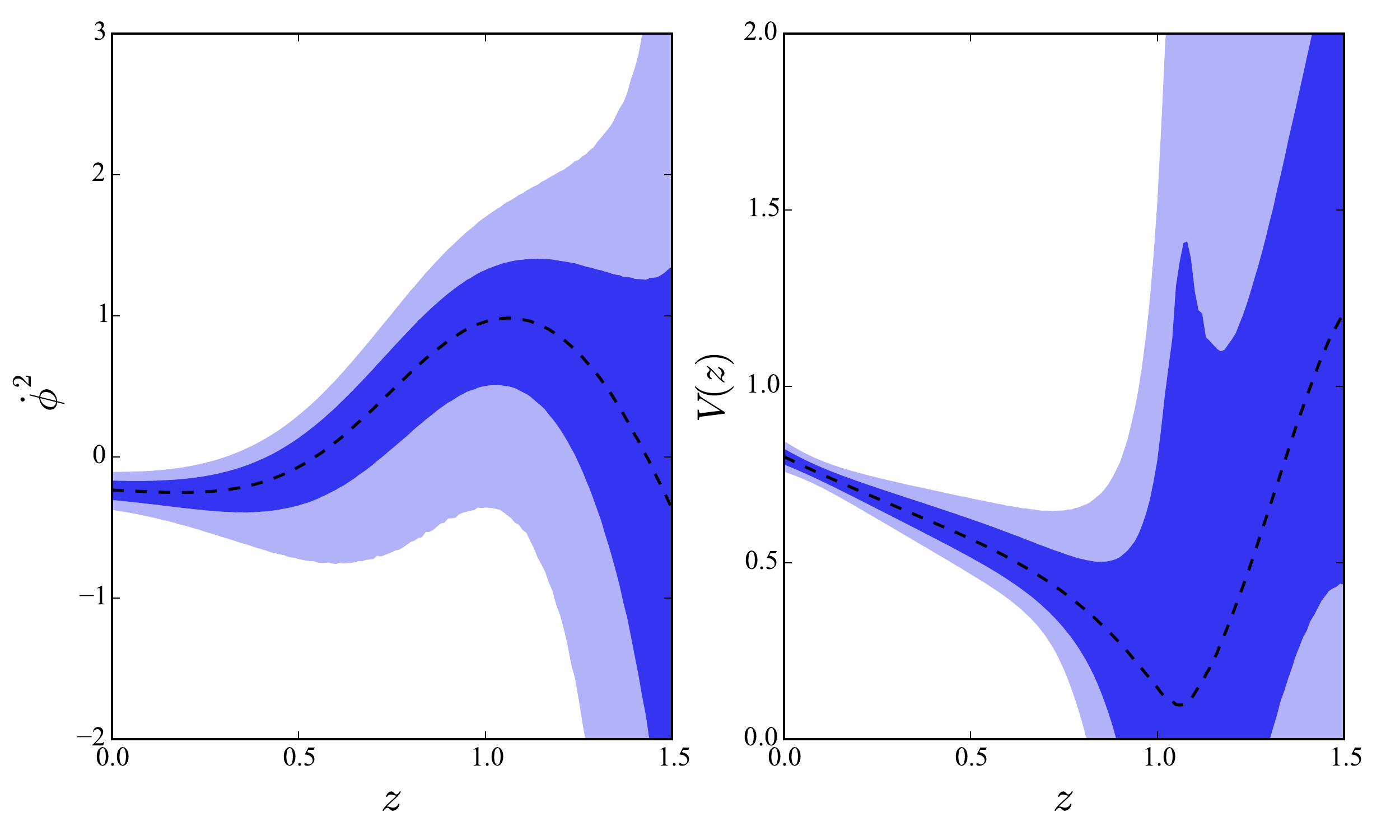}
    \caption{\label{JLAOHD_H07324_tachyon} GP reconstruction in the tachyon field for JLA and $H(z)$ data with Hubble constant $H_0=73.24 \pm 1.74$ km s$^{-1}$Mpc$^{-1}$. }
\end{figure}

\begin{figure}
\centering
    \includegraphics[width=7.5cm,height=5.5cm]{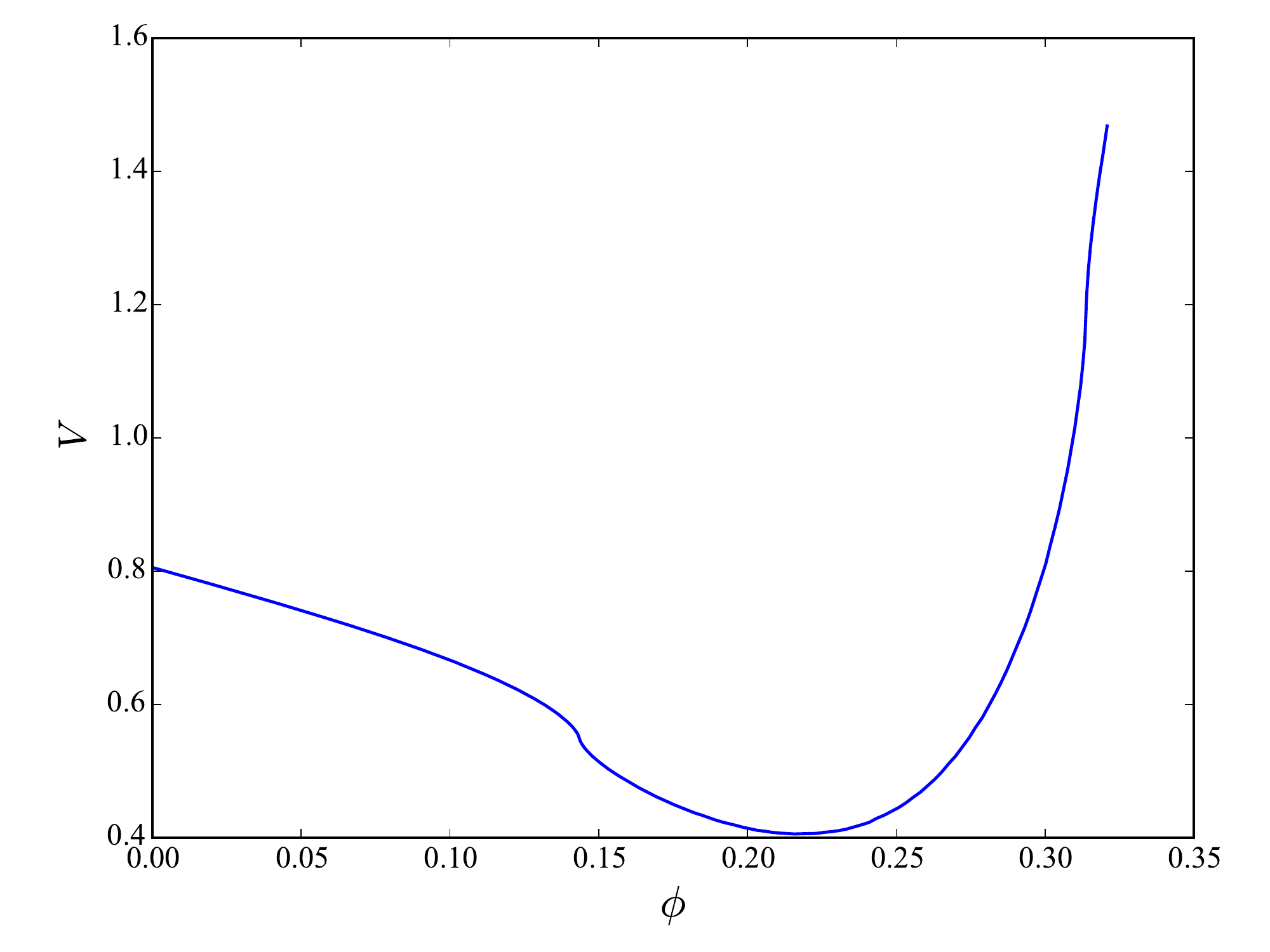}
    \caption{\label{VF7324_JLAOHD} The potential as a function of scalar field by their mean values for JLA and $H(z)$ data with Hubble constant $H_0=73.24 \pm 1.74$ km s$^{-1}$Mpc$^{-1}$. }
\end{figure}

\begin{figure}
\includegraphics[width=9.2cm,height=5cm]{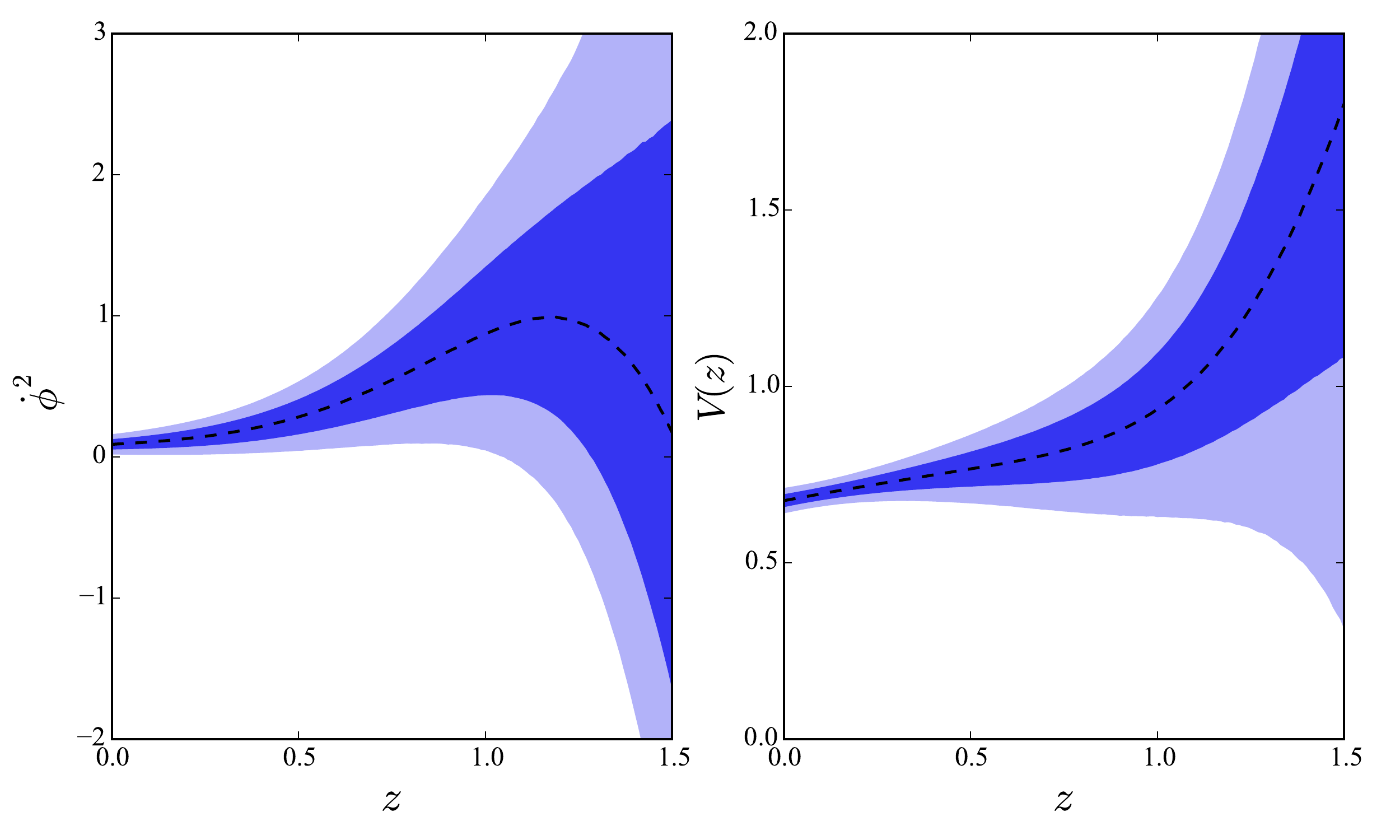}
    \caption{\label{H06774_quintess_JLAOHD} GP reconstruction in the \qu field for JLA and $H(z)$ data with Hubble constant $H_0=67.74 \pm 0.46$ km s$^{-1}$Mpc$^{-1}$. }
\end{figure}

To test the influence of Hubble constant on corresponding reconstructions, we report the results in two subsections.

\subsubsection{$H_0=73.24 \pm 1.74$ km s$^{-1}$Mpc$^{-1}$}
\label{result_H07324}

In Figs. \ref{JLAOHD_H07324_quintess} and \ref{JLAOHD_H07324_tachyon}, we plot the derivative of scalar field $\dot{\phi}^2$ and potential $V$ in the \qu field and tachyon field with $H_0=73.24 \pm 1.74$ km s$^{-1}$Mpc$^{-1}$. Theoretically, the function $\dot{\phi}^2$ should be $\dot{\phi}^2 \geq 0$. However, the figures show that it is negative at low redshift. At medium redshift, it turns to positive; and turns to negative again at high redshift. For the potential, $V$ decreases first and then increases at redshift $z \sim 1.0$. The initial value is $V_0 = 0.80$. Comparing the \qu field with tachyon field, they are similar. In Ref. \cite{calcagni2006tachyon}, the authors also found that tachyon models have quite similar phenomenology to canonical \qu models. Importantly, the ambiguous $\dot{\phi}^2$ indicates that the single \qu field, phantom field or tachyon field are all difficult to be favored by the data. Therefore, we cannot depict the function $V(\phi)$ using a single field. However, because the function $\dot{\phi}^2$  in \qu field and phantom field are opposite, we can also understand that it keeps switching between the two fields. So, the quintom field proposed by  \citet{feng2005dark} may be a better building on the scalar field. Now we treat the GP reconstruction in Fig. \ref{JLAOHD_H07324_quintess} as a quintom field, a combination of \qu and phantom field. Using the mean values of $\phi$ and $V$ in \qu field, we plot the potential as a function of field $\phi$ in Fig. \ref{VF7324_JLAOHD}, and find that it is a trough-form function. We fit this reconstruction with high $R$-square $= 0.9981$, and find that it obeys a 4-order Gaussian function, $V(\phi_q, \phi_p) = 0.3394   e^{- (\frac{\phi_q - 0.3203}{0.008478})^2 } + 0.19  e^{- (\frac{\phi_q + 0.01802}{0.04602})^2 }
+ 0.7125   e^{- (\frac{\phi_p - 0.05272}{0.1571})^2 } + 1.317 \times 10^{13} e^{- (\frac{\phi_p - 3.606}{0.5986})^2 }$, where $\phi_q$ and $\phi_p$ are the \qu field and phantom field, respectively. The fitted potential indicates that each potential should satisfy a double Gaussian function. We should reiterate that this fit is performed via their mean values. In the past, many parameterizations, such as the power-law, single exponential, etc. were proposed. The fitted potential can provide a reference. Moreover, we also emphasize that the $\dot{\phi}^2$ reconstruction within 68\% confidence level implies that the scalar field should be a quintom field --- a transformation between \qu field and phantom field.

\begin{figure}
\includegraphics[width=9.2cm,height=5cm]{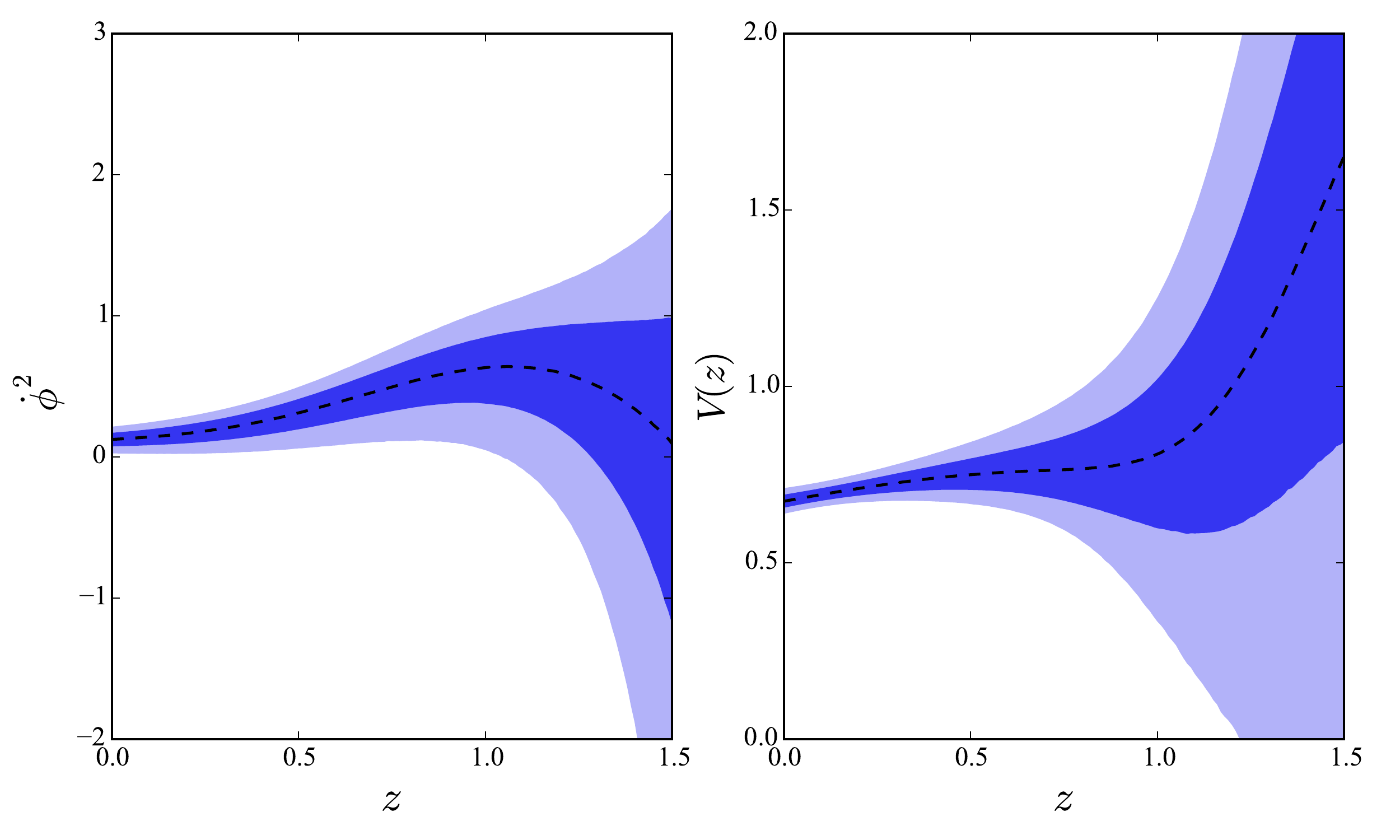}
    \caption{\label{H06774_tachyon_JLAOHD} GP reconstruction in the tachyon field for JLA and $H(z)$ data with Hubble constant $H_0=67.74 \pm 0.46$ km s$^{-1}$Mpc$^{-1}$. }
\end{figure}

\begin{figure}
\centering
    \includegraphics[width=7.5cm,height=5.5cm]{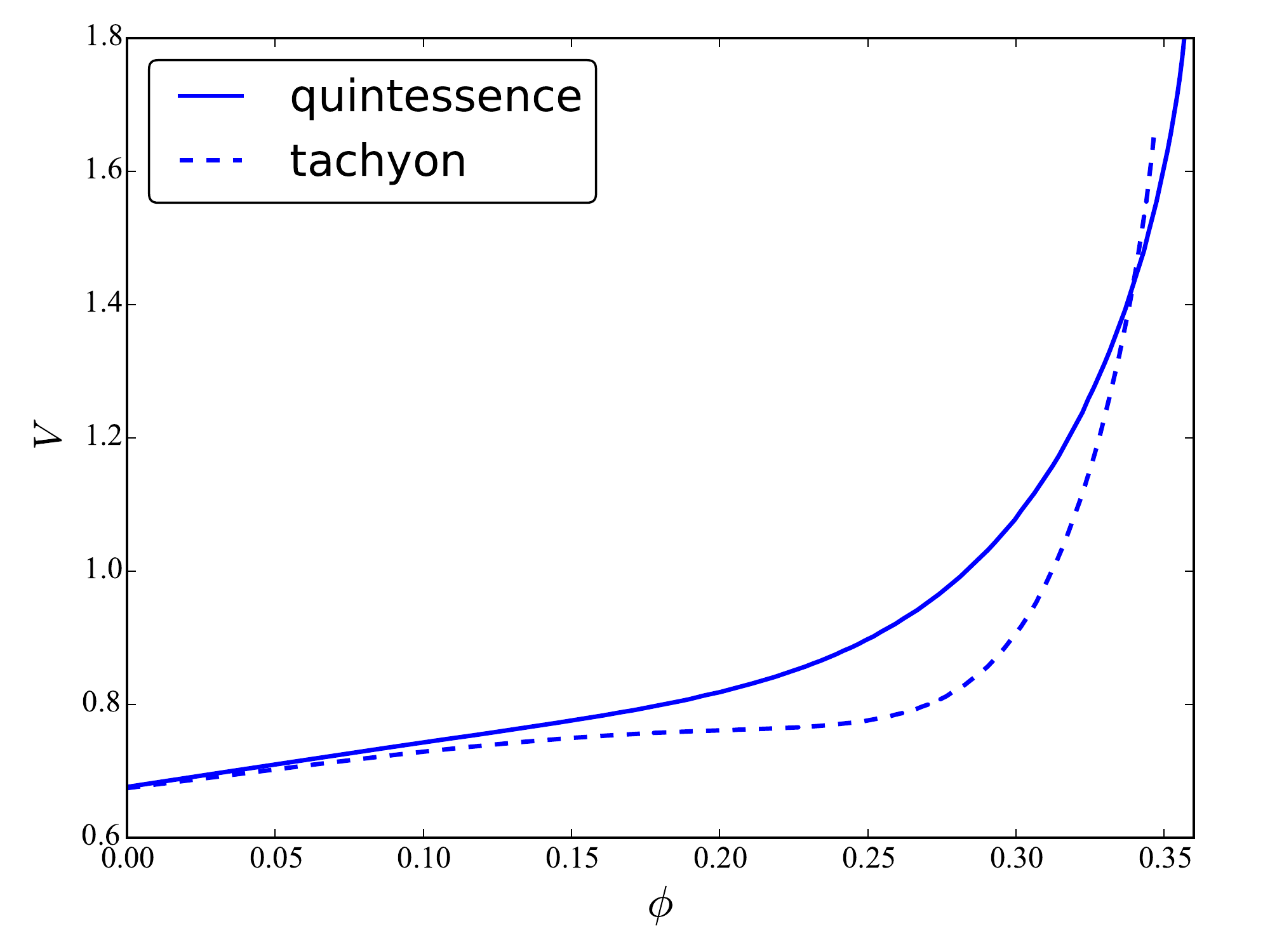}
    \caption{\label{H06774_VF_JLAOHD} The potential as a function of scalar field in different fields for JLA and $H(z)$ data with Hubble constant $H_0=67.74 \pm 0.46$ km s$^{-1}$Mpc$^{-1}$. }
\end{figure}

\subsubsection{$H_0=67.74 \pm 0.46$ km s$^{-1}$Mpc$^{-1}$}
\label{result_H06774}

In this prior, the JLA and $H(z)$ data present a slightly different reconstruction on these scalar fields.

In Fig. \ref{H06774_quintess_JLAOHD}, we plot the reconstructions in \qu field. Firstly, compared with the reconstructions in Fig.  \ref{JLAOHD_H07324_quintess}, the function  $\dot{\phi}^2$ also increases first and then decreases with the increasing redshift. Secondly, we find that mean values of the function $\dot{\phi}^2 >0$, which means that the \qu scalar field is favored to a certain degree. This situation is different from above reconstruction. However, what we should pay special attention to is that it still cannot prevent the derivative $\dot{\phi}^2 < 0$  at higher redshift within 68\% confidence level. That is, considering the uncertainties of $\dot{\phi}^2$, the quintom field is still a favourite model.  This result is consistent with the reconstruction in $H_0=73.24 \pm 1.74$ km s$^{-1}$Mpc$^{-1}$.  Last, the reconstruction of $V(z)$ shows that the data present an increasing potential. Especially for redshift $z \gtrsim 1$, it increases sharply.  At redshift $z=0$, we have a model-independent estimation $V_0 = 0.68$, which is similar as above reconstruction.

In Fig.  \ref{H06774_tachyon_JLAOHD}, we plot the reconstructions in tachyon field. We find that the data present a similar reconstruction as the \qu field. That is, mean values of the function $\dot{\phi}^2$ are also positive; and the potential $V(z)$  is also an increasing function. However, within 68\% confidence level, the function $\dot{\phi}^2 < 0$ is still supported by the data. So, the tachyon field cannot be convincingly favored by the data.

In Fig. \ref{H06774_VF_JLAOHD}, we plot the function $V(\phi)$ using their mean values. From this figure, we find that the data present a same initial value of potential $V_0 = 0.68$ in both \qu field and tachyon field. With the increase of $\phi$ field, potentials first increase slowly with a similar rate. However, with the continuous increase of $\phi$ field, potential in the \qu field increases faster than that in the tachyon field. Therefore, they may reflect a slightly different scalar field model.

Now, we fit the function $V(\phi)$ in different fields, and present the list in table \ref{tab: V_phi}. For the \qu field, the mean values favor a double function, such as $V(\phi) = 0.677 e^{0.8973 \phi} + 7.077 \times 10^{-5} e^{26.24 \phi}$. As pointed out in above subsection, the potential should be a double exponential function or Gaussian function.

\begin{table}
\caption{\label{tab: V_phi} Function $V(\phi)$ obtained by their mean values in different scalar fields for different observational data. }
\begin{tabular}{ll}
\hline
\hline
JLA+$H(z)$:   & quintom               \\
\scriptsize{$(H_0=73.24 \pm 1.74 )$}  \\
\hline
Gaussian   :       & $V(\phi_q, \phi_p) = 0.3394   e^{- (\frac{\phi_q - 0.3203}{0.008478})^2 } + 0.19  e^{- (\frac{\phi_q + 0.01802}{0.04602})^2 }$  \\
                   & \qquad \quad $  + 0.7125   e^{- (\frac{\phi_p - 0.05272}{0.1571})^2 } + 1.317 \times 10^{13} e^{- (\frac{\phi_p - 3.606}{0.5986})^2 }$  \\
\hline
\nonumber \\

JLA+$H(z)$:   & \qu               \\
\scriptsize{$(H_0=67.74 \pm 0.46)$} \\
\hline
Exponential:               & $V(\phi) = 0.677 e^{0.8973 \phi} + 7.077 \times 10^{-5} e^{26.24 \phi}$      \\
Gaussian   :       & $V(\phi) = 7.17 \times 10^{15} e^{- (\frac{\phi - 3.168}{0.4641})^2 } + 1.010 \times 10^{3} e^{- (\frac{\phi - 15.96}{5.905})^2 }$      \\
\hline
\nonumber \\

JLA+$H(z)$:   & tachyon                  \\
\scriptsize{$(H_0=67.74 \pm 0.46)$}   \\
\hline
Exponential:               & $V(\phi) = 0.6927 e^{0.4208 \phi} + 5.262 \times 10^{-7} e^{41.24 \phi}$      \\
Gaussian   :       & $V(\phi) = 6.018 \times 10^{4} e^{- (\frac{\phi - 0.9633}{0.1851})^2 } + 0.7578 e^{- (\frac{\phi - 0.2208}{0.63})^2 }$      \\
\hline
\nonumber \\

RSD:   & quintom                 \\
\hline
Gaussian   :       & $V(\phi_q, \phi_p) = 0.06178   e^{- (\frac{\phi_q - 0.2588}{0.0771})^2 } + 0.4105 e^{- (\frac{\phi_q - 0.3806}{0.1371})^2 } $  \\
                   & \qquad \quad $  + 0.9255   e^{- (\frac{\phi_p - 0.1266}{0.2122})^2 } -9.546 \times 10^{11} e^{- (\frac{\phi_p - 17.87}{3.233})^2 } $  \\

\hline
\hline
\end{tabular}
\end{table}

\subsection{Reconstruction from the RSD data}
\label{resultRSD}

\begin{figure}
\includegraphics[width=9.2cm,height=5cm]{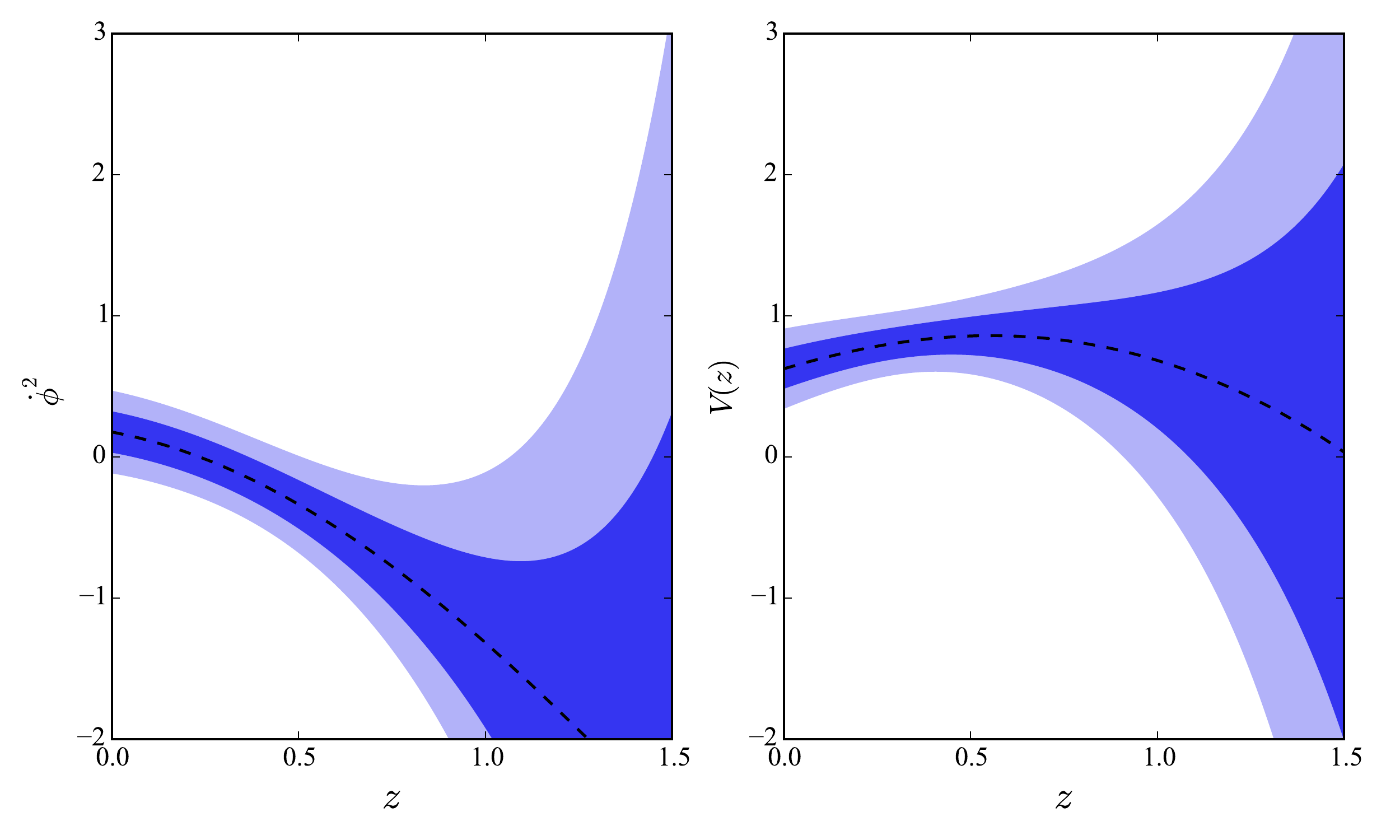}
    \caption{\label{quintess_RSD}  GP reconstruction in the \qu field for RSD data.}
\end{figure}

In Fig. \ref{quintess_RSD}, we plot the reconstruction in \qu field. We find that these results are different from the reconstruction for JLA and $H(z)$ data. Firstly, function $\dot{\phi}^2$ from the background data increases first and then decreases with the increasing redshift. While for the RSD data, the function $\dot{\phi}^2$ slows down sharply from positive to negative, which indicates a direct transformation from \qu field to phantom field. Different from the Fig. \ref{H06774_quintess_JLAOHD}, potential $V$ from the RSD data is a decreasing function. Therefore, we think that the RSD data may present a quite different scalar field model. Secondly, the initial value of potential for two different types of data are similar. That is, the RSD data present $V_0 = 0.62$, which is similar as  $V_0 = 0.68$ from the background data.

\begin{figure}
\includegraphics[width=9.2cm,height=5cm]{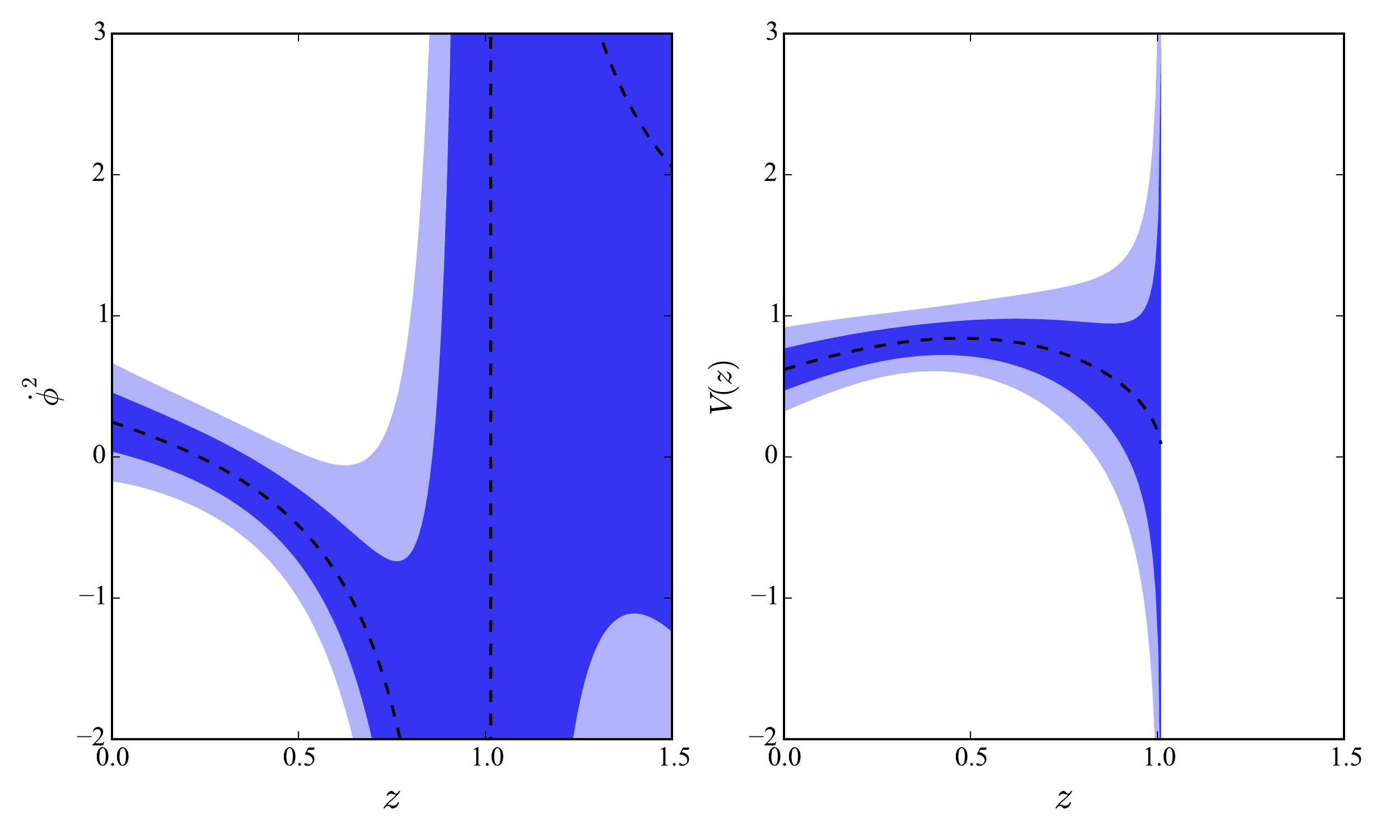}
    \caption{\label{tachyon_RSD}  GP reconstruction in the tachyon field for RSD data. }
\end{figure}

In Fig. \ref{tachyon_RSD}, we plot the reconstruction in tachyon field. At low redshift, the function $\dot{\phi}^2$ also transfers from positive to negative, which is similar as Fig. \ref{quintess_RSD} for \qu field. However, it has a shock change at redshift $z \sim 1.0$, and then changes to positive. Because it cannot ensure the function $\dot{\phi}^2 > 0$, we think that the tachyon field cannot be convincingly favored by the data. This determination is same as the results for JLA and $H(z)$ data. For the potential $V$, it is absent at redshift $z \gtrsim 1.3$. This is because the potential $V$ in Eq. \eqref{tachyon:phi2t} at high redshift is invalid. Combining with the singularity in function $\dot{\phi}^2$, we think that the tachyon field is at a disadvantage to describe the cosmic evolution.

\begin{figure}
\centering
    \includegraphics[width=7.5cm,height=5.5cm]{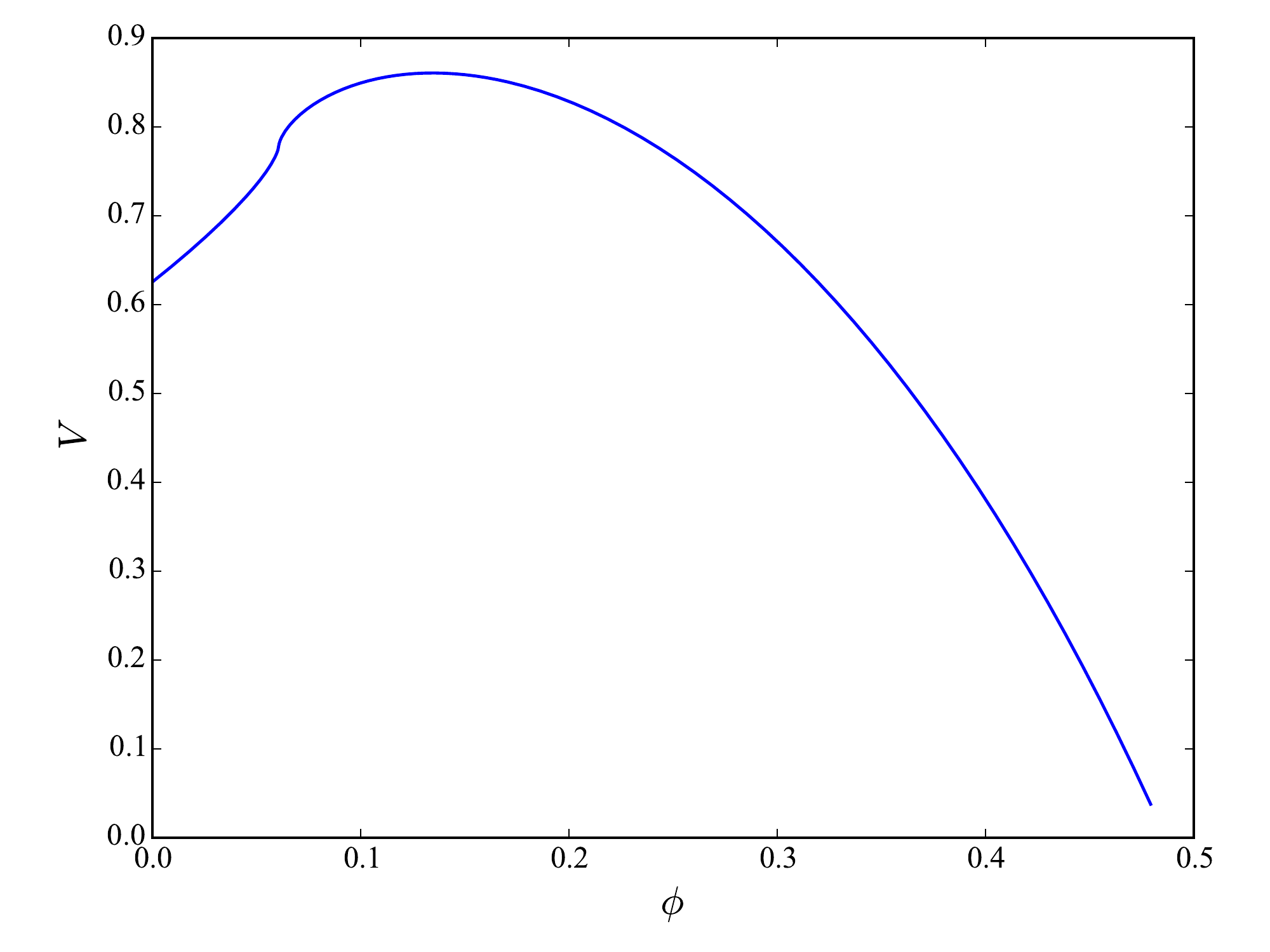}
    \caption{\label{VF_RSD} The potential as a function of scalar field in quintom field for RSD data. }
\end{figure}

In Fig. \ref{VF_RSD}, we plot the function $V(\phi)$ using their mean values. As pointed out above, the function $\dot{\phi}^2$ in Fig. \ref{quintess_RSD} cannot fulfill $\dot{\phi}^2 > 0$ at all redshift. It transfers from positive to negative at redshift $z \sim 0.24$. To describe the reconstructed scalar field, we understand it as a quintom field, which changes from \qu field to phantom field. From the picture, we find that it is also a complicated model similar as the Fig. \ref{VF7324_JLAOHD}. However, it seems to be reciprocal to Fig. \ref{VF7324_JLAOHD}. From the table \ref{tab: V_phi}, we find that $V$ in each scalar field should satisfy a double exponential or double Gaussian function, which is consistent with the reconstruction for background data.

\section{Conclusion and discussion}
\label{conclusion}

In this paper, we carry out a model-independent test on the scalar field to explore the source of dark energy, using the Gaussian processes approach. The observational data we use are supernova data, $H(z)$ parameter and growth rate data.

In past several years, the scalar field has been studied via many parameterizations. Focus of attention was which template is the best dynamical description of dark energy. Work in this paper not only presents a template-free analysis, but also reconfirm that the dark energy should be dynamical.

From the background data, we find that it does not favor a single \qu field or phantom field or tachyon field in the prior of $H_0=73.24 \pm 1.74$ km s$^{-1}$Mpc$^{-1}$. Within 68\% confidence level, it favors a quintom field, which is a transformation between \qu  field and phantom field. Using their mean values, the fitted potential $V(\phi)$  is a 4-order Gaussian function, which indicates that potential in each field is a double exponential or Gaussian function. We also test the influence of Hubble constant, and find that $H_0$ has a notable influence on the reconstruction. In the prior of $H_0=67.74 \pm 0.46$ km s$^{-1}$Mpc$^{-1}$, mean values of $\dot{\phi}^2$ favor the \qu field with a double exponential potential, such as $V(\phi) = 0.677 e^{0.8973 \phi} + 7.077 \times 10^{-5} e^{26.24 \phi}$. However, when considering their uncertainties, the reconstructions also favor the quintom field.

Our study also solves another puzzle. In previous work \cite{calcagni2006tachyon}, it was found that the  tachyon models have quite similar phenomenology to canonical \qu models. We find that they are similar at low redshft (or small  $\phi$). However, with the increase of values of $\phi$ field, they gradually split up, as shown in Fig. \ref{H06774_VF_JLAOHD}. Moreover, the background data and RSD data both reveal that the tachyon field is at a disadvantage to describe the cosmic acceleration.

From the RSD data, they also prefer to a quintom field. This determination is identical with above analysis from the background data. Moreover, mean values show that potential $V(\phi)$ in the quintom field is also a 4-order Gaussian function. In addition, the RSD data show a more remarkable difference between the tachyon field and \qu field, as displayed in Figs. \ref{quintess_RSD} and \ref{tachyon_RSD}.

Argument about the dynamics of dark energy has been going on around whether it evolved or not. Our analysis highly reveals that the dark energy should be dynamical, regardless of it is from the background data or perturbation data. Moreover, the corresponding reconstructions indicate that the data favor the quintom field with high significance. In recent work by Zhao et al. \cite{zhao2017dynamical}, they investigated the dark energy using the latest data including CMB temperature and polarization anisotropy spectra, supernova, BAO from the clustering of galaxies and from the Lyman-$\alpha$ forest, Hubble constant and $H(z)$.  They found that the dynamical dark energy can relieve the Hubble constant tension and is preferred at a 3.5$\sigma$ significance level. Moreover, the upcoming dark energy survey DESI++ would be able to provide a decisive Bayesian evidence. In our future work, we also would like to invest more observational data on the study of scalar field, to make a clearer analysis on cosmic dynamics.

Another point we should emphasize is the importance of Hubble constant. We note that it has a notable influence on the background data reconstruction. Its precise measurement can discriminate the scalar field is \qu field or quintom field. The tension in $H_0$ has aroused great concern. Some Refs. \cite{freedman2017cosmology,mortsell2018does} think that it may be a signature of new physics. Till now, its measurement window has been opened from traditional Cepheids, Tip of the Red Giant Branch, Type Ia Supernovae, Surface Brightness Fluctuations, Masers, and Gravitational Lens Time Delays, to fashionable Gravitational-wave \cite{schutz1986determining}. The detection of GW170817-a strong signal
from the merger of a binary neutron-star system and the identification of its host galaxy has obtained a completely independent and consistent determination with existing measurements \cite{ligo2017gravitational}. Moreover, it also can be measured with neutron star black hole mergers from advanced LIGO and Virgo \cite{PhysRevLett.121.021303}. The future multi-messenger astronomy will enable the $H_0$ and cosmic dynamics to be constrained to high precision.

\section*{Acknowledgments}

M.-J. Zhang  would like to thank Shulei Ni, De-Liang Wu, Hua Zhai and Xiao Wang for valuable discussion. H. Li is supported by the Youth Innovation Promotion Association Project of CAS. M.-J. Zhang is funded by China Postdoctoral Science Foundation under grant No. 2015M581173. The research is also supported in part by NSFC under Grant Nos. 11653001, Pilot B Project of CAS (No. XDB23020000) and Sino US Cooperation Project of Ministry of Science and Technology (No. 2016YFE0104700).

\bibliography{scalar}
\end{document}